\documentclass[12pt]{iopart} 

\usepackage{amsfonts}
\usepackage{amssymb}
\usepackage{graphicx}
\usepackage{subfig} 
\usepackage{bm}
\usepackage{latexsym}

\def\a{\alpha}
\def\l{\ell}

\def\bl{{\bar\ell}}
\def\av#1{\left\langle#1\right\rangle}

\def\comment#1{}
\newcommand{\be}{\begin{equation}}
\newcommand{\ee}{\end{equation}}
\newcommand{\bea}{\begin{eqnarray}}
\newcommand{\eea}{\end{eqnarray}}

\begin{document}

\title{Small-scale behaviour in
deterministic reaction models}
\author{Paolo Politi$^{1}$ and Daniel ben-Avraham$^{2,3}$}
\address{$^1$ Istituto dei Sistemi Complessi, Consiglio Nazionale
delle Ricerche, Via Madonna del Piano 10, 50019 Sesto Fiorentino, Italy}
\address{$^2$ Physics Department, Clarkson University, Potsdam, NY 13699-5820, USA}
\address{$^3$ Department of Mathematics, Clarkson University, Potsdam, NY 13699-5815, USA}
\eads{
\mailto{paolo.politi@isc.cnr.it},
\mailto{benavraham@clarkson.edu}
} 
\begin{abstract}
In a recent paper published in this journal [{\bf 42} (2009) 495004]
we studied a one-dimensional particles system where nearest particles 
attract with a force inversely proportional to a power $\a$ of their distance 
and coalesce upon encounter.  
Numerics yielded a distribution function $h(z)$ for the gap between
neighbouring particles, with $h(z)\sim z^{\beta(\a)}$ for small $z$
and $\beta(\a)> \alpha $. 
We can now prove analytically that in the strict  
limit of $z\to 0$, $\beta=\a$ for $\a>0$, corresponding to the mean-field
result, and we compute the length scale where mean-field breaks down. 
More generally, in that same limit correlations are negligible 
for any similar reaction model where attractive forces diverge
with  vanishing distance.
The actual meaning of the measured exponent $\beta(\a)$ remains an open question.
\end{abstract}


\maketitle  

\section{Introduction}

In a recent publication~\cite{ourJPA} we have 
studied an infinite system of particles on the line, located at $\{x_i(t)\}_{i=-\infty}^{\infty}$, when
nearest particles attract one another with a force inversely proportional
to a power $\a$ of the distance and coalesce upon encounter.
In the overdamped limit, we can write
\begin{equation}
\label{dx/dt}
\frac{dx_i}{dt}=\frac{A}{\a}\left(\frac{1}{(x_{i+1}-x_i)^\a}-\frac{1}{(x_i-x_{i-1})^\a}\right)
\end{equation}
and the gaps between particles, $\l_i=x_{i+1}-x_{i}$, obey
\begin{equation}
\label{dl/dt}
\frac{d\l_i}{dt}=\frac{A}{\a}\left(\frac{1}{\l_{i-1}^\a}-\frac{2}{\l_i^\a}+\frac{1}{\l_{i+1}^\a}\right)\,.
\end{equation}
Reacting particles systems such as this, evolving deterministically, occur often enough to
merit further study~\cite{PhysD,kawasaki,krapivsky,bray,krug,redner,derrida}.  The case of $\a=1$
occurs in the study of crystal growth~\cite{PhysD}.  For a more detailed discussion  see Ref.~\cite{ourJPA}.

In the following we will consider the case of  $\a>0$,
which corresponds to attractive forces that diverge as the gap
between nearest particles vanishes. The system is unstable and 
neigbouring particles tend to attract and
coalesce, which leads to a reduction in the number of particles,
i.e., to a coarsening process.

\begin{figure}[ht]
\vspace*{0.cm}
\includegraphics*[width=0.8\textwidth]{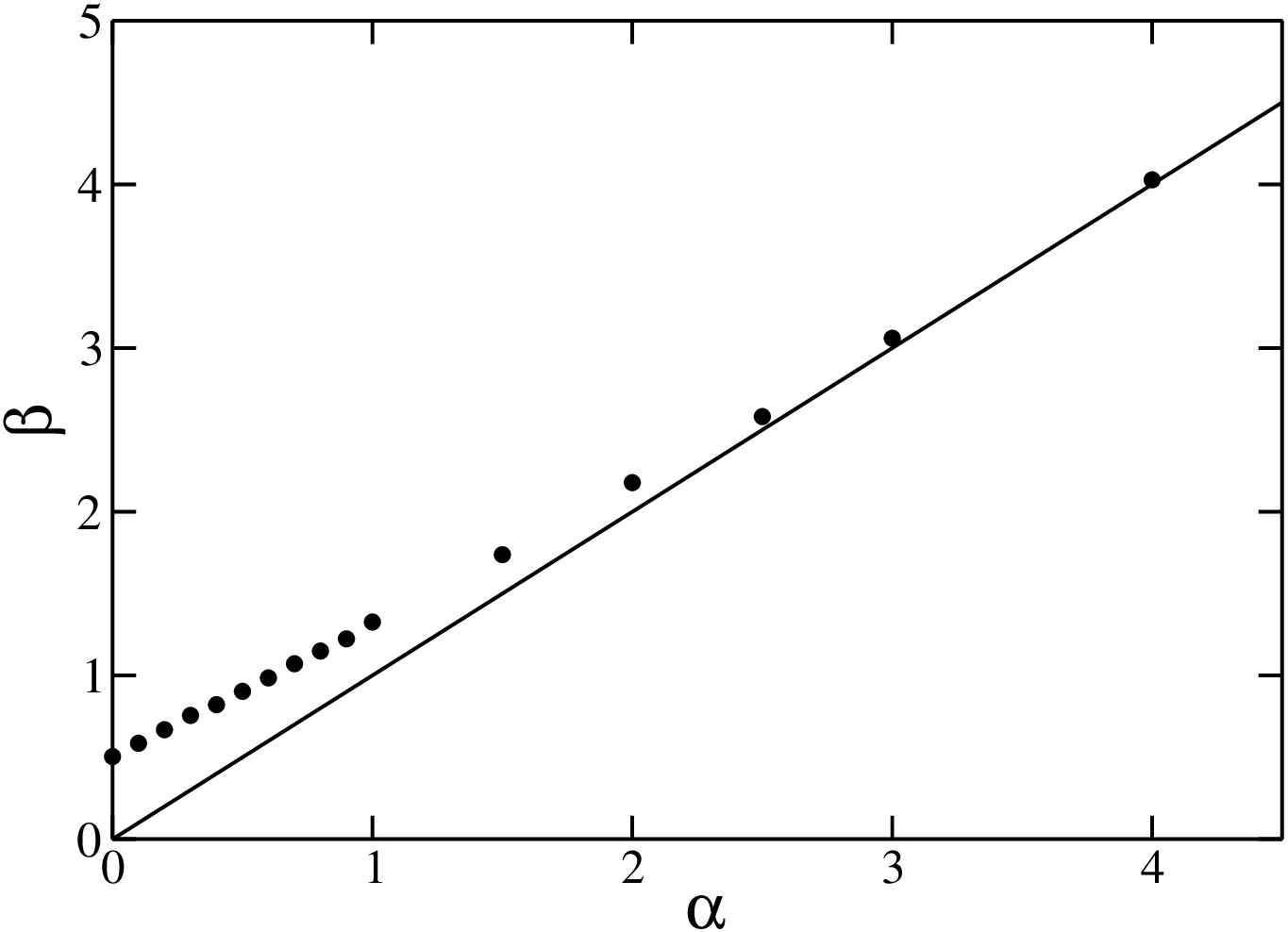}
\caption{The exponent $\beta$, that describes the small gap behaviour of the gap distribution 
function, $h(z)\sim z^{\beta}$, as a function of the force exponent $\a$, in our deterministic 
reaction model.  The solid line indicates the strict $z\to0$ limit $\beta=\alpha$ of Eq.~(\ref{b=a}).
After Fig.~1 of Ref.~\cite{ourJPA}.}
\label{fig.beta}
\end{figure}

Dimensional analysis, as well as a scaling hypothesis
applied to the pertinent Fokker-Plank equations, yield the coarsening law~\cite{ourJPA}
for the average distance $\bl$ between nearest neighbour particles,
\be
\label{coarsening}
\bl(t) = [2A(1+\a)t]^n\,,\qquad n={1/(1+\a)}\,.
\ee
This  firm theoretical result is also strongly supported by unambiguous numerical evidence. 
In contrast, the distribution function $h(z)$, for the reduced
gap $z=\l/\bl$ between particles, has proved to be more challenging,
both numerically and analytically. In~\cite{ourJPA}
we have presented numerical simulation results supporting a power-law behavior for small $z$,
\be
h(z) \sim z^{\beta(\a)}\,,
\ee
where $\beta(\alpha)>\alpha$: $\beta(0)\approx1/2$, and $\beta(\alpha)\to\alpha$ 
as $\alpha\to\infty$ (Fig.~\ref{fig.beta}).

Analytically, we were unable to do better than mean-field, which predicts
$\beta=\a$.  However, qualitative considerations had
suggested that $\beta$ might be larger than $\a$, in the presence
of sufficiently strong anti-correlations between adjacent gaps.

Below, we show analytically that the exponent $\beta$ 
and the coarsening exponent $n$ are related to one another.
It follows from this relationship that in the strict limit of $z\to0$, $\beta=\alpha$.
The remainder of the paper is devoted to the question of why numerics yield
a different value for $\beta$.

\section{Relation between the exponents $\beta$ and $n$}

Assuming that $\alpha>0$, Eq.~(\ref{dl/dt}) tells us that once the gap $\l_i$ is
small enough its own size dominates its rate of shrinking and ultimately this
rate becomes independent of the neighbouring gaps, $\l_{i\pm1}$.
Note that $h(z)\to0$ as $z\to0$, so the probability that either of the $\l_{i\pm1}$ be
also small can be safely neglected.
Thus, in this limit,
\be
\frac{d\l_i}{dt}=-\frac{2A}{\a\l_i^\a}\,,
\label{limit_dl/dt}
\ee
whose integrations gives
\be
\l_i^{\a+1}(0)-\l_i^{\a+1}(t)= 2A\left({\a+1\over\a}\right) t.
\ee
If $\tau$ is the time that it takes for a small gap of size $\l_i(0)=\delta$ to close up,
$\l_i(\tau)=0$ 
(resulting in a coalescence event), we get
\be
\delta^{\a+1} = 2A\left(\frac{\a+1}{\a}\right) \tau\,.
\label{lc}
\ee

Suppose that at time $t$ we have $N$ particles.  Between $t$ and $t+\tau$ 
all gaps smaller than $\delta$ would coalesce, so the fraction of coalescence
events during the time $\tau$ is equal to the fraction of intervals smaller than $\delta$,
\bea
{dN\over N} &=& -\int_0^{\delta/\bl} dz\, h(z)
= - {B\over\beta+1}\left({\delta\over\bl}\right)^{\beta+1} \nonumber\\
&=& - {B\over\beta+1} {1\over \bl^{\beta+1}} 
\left[ 2A\left({\a+1\over\a}\right) \tau\right]^{\beta+1\over\a+1} \, ,
\label{dN1}
\eea
where we have assumed the small-$z$ behaviour
$h(z)=B z^\beta$ and we have used Eq.~(\ref{lc}).
On the other hand, since $\bl=L/N$, where $L$ is the constant, total length 
of the system, $d\bl=-\bl dN/N$, or $dN/N=-d\bl/\bl$. 
Differentiating Eq.~(\ref{coarsening}) for an infinitesimal time $dt=\tau$, we get
\be
{dN\over N} = -{2A\over\bl^{\a+1}}\tau \, .
\label{dN2}
\ee
We can now equate the right hand sides of Eqs.(\ref{dN1}) and (\ref{dN2}),
\be
{B\over\beta+1} {1\over \bl^{\beta+1}} 
\left[ 2A\left({\a+1\over\a}\right) \tau\right]^{\beta+1\over\a+1}
= {2A\over\bl^{\a+1}}\tau\,,
\ee
which must be equal for any small $\tau$. This implies
\be
\label{b=a}
\beta = \a\quad{\rm and}\quad
B = \a\,.
\ee

It is worth mentioning
that the exponents $n$ and $\beta$ can be related, in principle, following the same
procedure as above, also when
the attractive force does not diverge with vanishing gap (as for,
e.g., an exponential force, $f(\l)\sim e^{-\l}$), or when the force is
repulsive (our model, with $\a<0$). However,
in such cases adjacent gaps influence the outcome and correlations are important 
even in the limit of
vanishing $z$, which precludes a simple derivation
of $\delta(\tau)$.

\section{Correlation effects: discussion and conclusions}

Numerically, the observed small-gap exponent,
$\beta$, seems to be larger than $\alpha$, the value predicted from the strict 
limit of $z\to0$ (Fig.~\ref{fig.beta}).  How small need $z$ be
for the result $\beta=\alpha$ to hold, even approximately?

Assuming that the small gap $\delta$ is surrounded by gaps of typical length $\bl$,
according to Eq.~(\ref{dl/dt}) one can neglect their influence when
$1/\delta^{\a}\gg {1/ \bl^\a}$, or
\be
\left( \frac{\delta}{\bl}\right)^\a <\varepsilon\,,
\ee
where $\varepsilon$ is  some small positive  number representing our error tolerance. 
We can expect that correlations with adjacent intervals are negligible at (reduced) distances 
$z<\varepsilon^{1/\a}$.
This is consistent with the findings of Fig~(\ref{fig.beta}) that the error is larger for smaller $\alpha$. 

We can make a more accurate assessment of the influence of correlations from neighbouring gaps.
Let $h_2(z_1,z_2)$ be the joint probability density for adjacent gaps of reduced lengths $z_1$ and $z_2$,
and denote by
\be
\av{1\over z_1^\a}_{z} =
\int_0^\infty dz_1 {1\over z_1^\a} {h_2(z_1,z)\over h(z)}
\ee
the conditional average of $z_1^{-\a}$, given that the adjacent interval has length $z_2=z$.
This conditional average obeys the {\it exact} expression (Eq.~(30), Ref.~\cite{ourJPA})
\begin{equation}
\label{h}
2\a h(z)+\a zh'(z)=\left[\left(\av{\frac{1}{z_1^\a}}_z-\frac{1}{z^\a}\right)h(z)\right]'\,,
\end{equation}
where the prime denotes differentiation with respect to $z$.
Numerically, one measures the exponent $\beta$ from the slope
of a log-log plot of $h(z)$ vs.~$z$.  Rearranging~(\ref{h}), we obtain for the {\em local}
slope:
\begin{equation}
\label{hratio}
g_\a(z)=
\frac{d(\ln h)}{d(\ln z)} = 
\frac{zh(z)'}{h(z)}=z\frac{-2\a+\a z^{-\a-1}+\av{z_1^{-\a}}'_z}%
{\a z+z^{-\a}-\av{z_1^{-\a}}_z}\,.
\end{equation}

Further progress depends on $\av{z_1^{-\a}}_z$.  At the simplest level,
mean-field says that $h_2(z_1,z_2)=h(z_1)h(z_2)$, leading to  a {\em constant}
value of the conditional average, which can be shown to be~\cite{ourJPA}
\be
\label{mf}
\left\langle {1\over z_1^\a}\right\rangle_z = 1+\a \,.
\ee
%
%
Another tractable possibility is to assume that adjacent gaps are perfectly anti-correlated:
as one interval grows the adjacent gap shrinks, and their total length
is fixed, $\l_1+\l_2=2\bl$, such that $h_2(z_1,z_2)=h(z_1)\delta(2-z_1-z_2)$.
This leads to
\be
\label{corr}
\left\langle {1\over z_1^\a}\right\rangle_z = {1\over (2-z)^\a} \,.
\ee

\begin{figure}[ht]
\vspace*{0.cm}
\includegraphics*[width=0.47\textwidth]{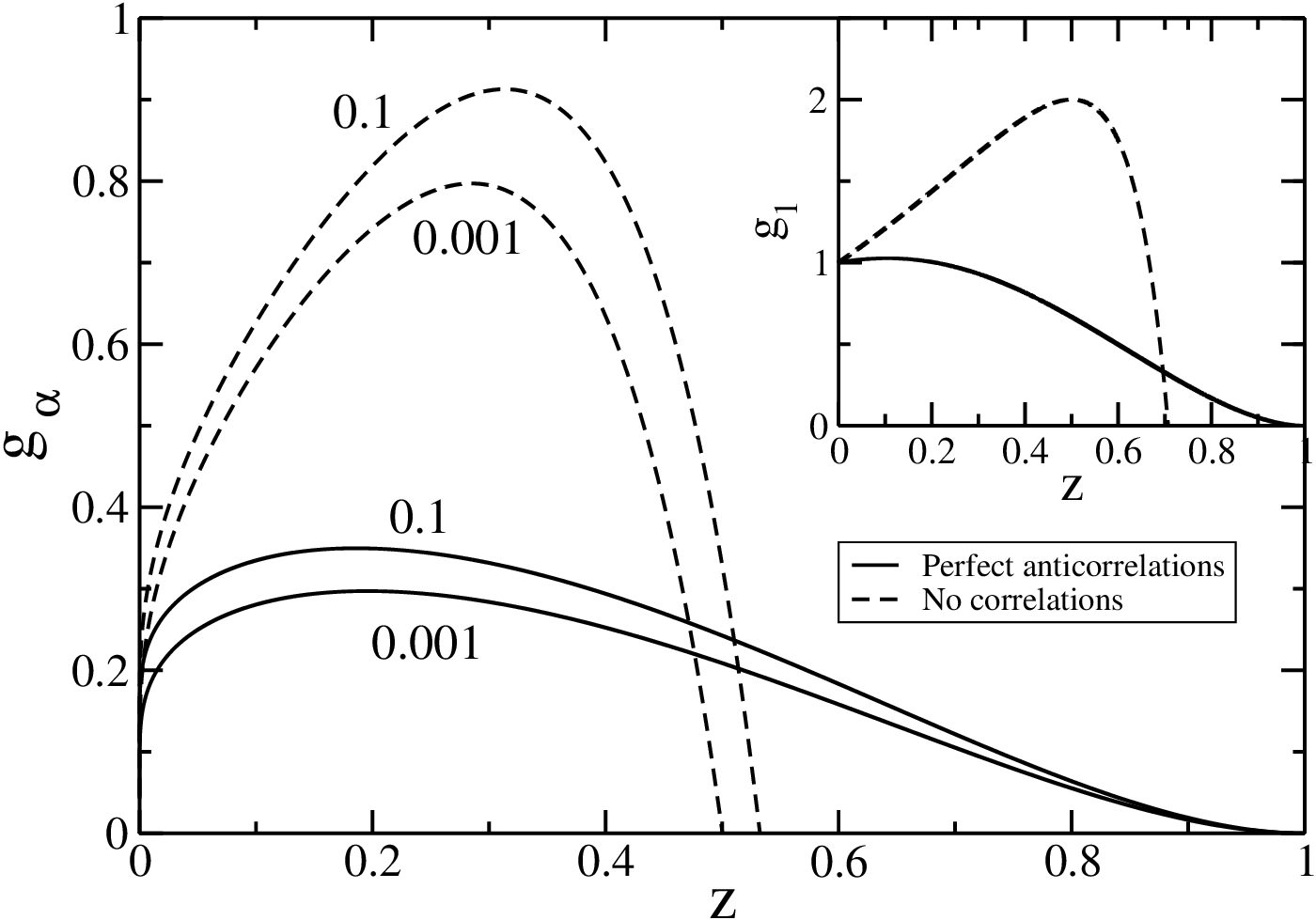}\hskip0.2in
\includegraphics*[width=0.47\textwidth]{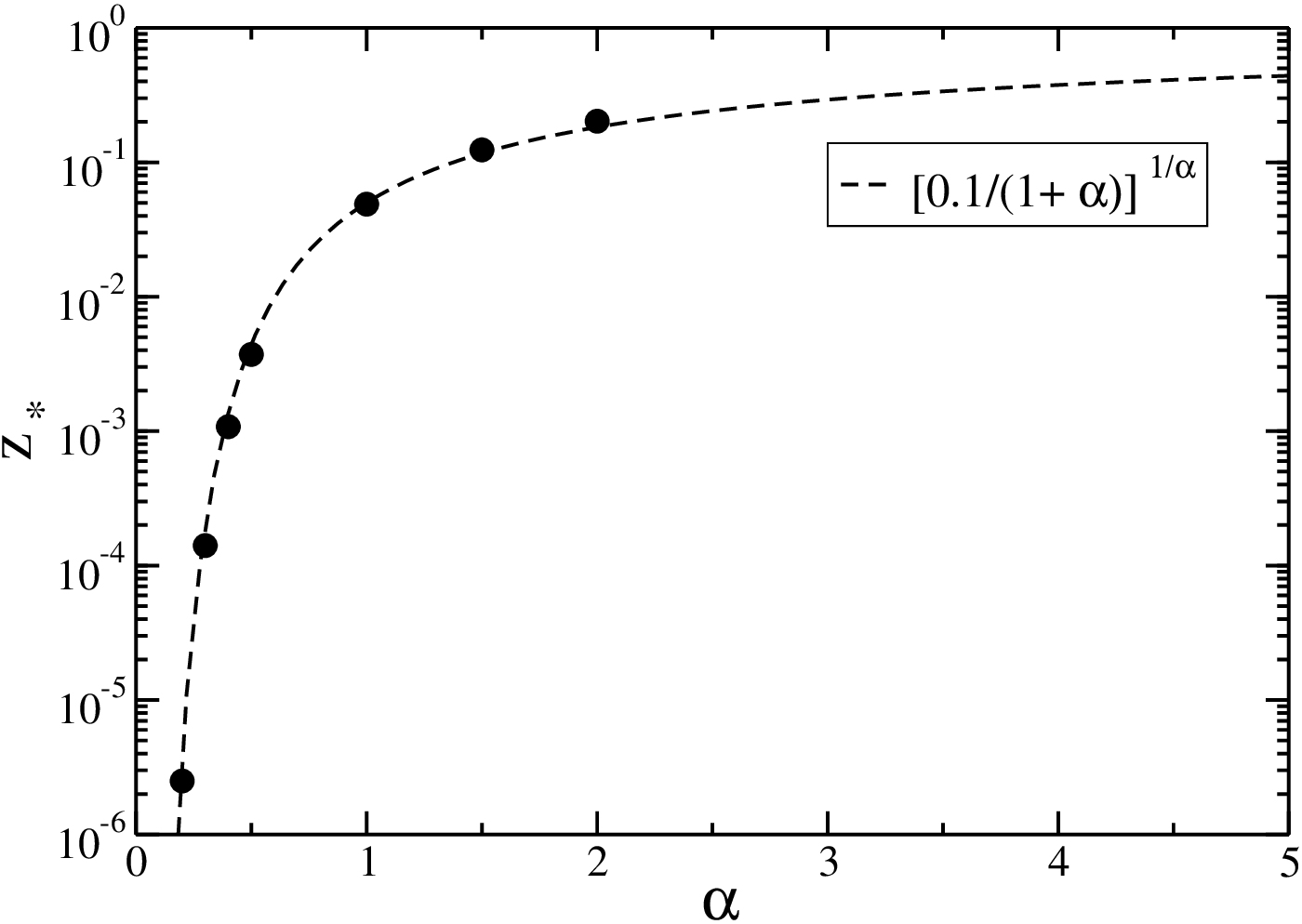}
\caption{(a)~Plot of the local slope $g_\a(z)$ for $\a=0.001$, $0.1$, and $1$ (inset), for
the mean-field case (dashed lines) and for perfect correlations (solid lines). (b)~The boundary $z_*$
within which the mean-field local slope deviates from $\a$ by less than $10\,\%$, as obtained
numerically from (\ref{hratio}) and (\ref{mf}) (symbols) compared to the first-order approximation
(\ref{zstar}) (dashed line).  Note the logarithmic scale.}
\label{fig.g}
\end{figure}

In Fig.~\ref{fig.g}a we plot the local slope $g(z)$ for the two assumptions, 
(\ref{mf}) and (\ref{corr}), for $\a=0.001$, $0.1$, and $1$ (inset). In all cases, 
it is easy to confirm analytically that $g(0)=\alpha$, in agreement with~(\ref{b=a}).
Also evident from the figure, is the fact that $g(z)$ rises sharply near the origin (with infinite
slope, for $\a<1$), which may explain how numerically one may observe an effective
exponent larger than $\alpha$.

 We can also use the expression for the local slope, (\ref{hratio}), in conjunction
 with~(\ref{mf}), to determine the condition for $\a\leq\beta\leq\a(1+\varepsilon)$.
 Solving to first order in $\varepsilon$ we find that $0<z<z_*$, with
\be
\label{zstar}
z_* \simeq \left( {\varepsilon\over1+\a}\right)^{1/\a}\,,
\ee
(and a similar expression for perfect anti-correlations).  Note the similarity
to the criterion for neglecting adjacent intervals, derived in the previous section.
In Fig.~\ref{fig.g}b we plot $z_*$ for $\varepsilon=0.1$, as a function of $\a$.  The conclusion
is that a change in $\beta$ as big as $10\,\%$ occurs, for most values of $\a$, within a very small
range of $z$.  For example, $z_*<10^{-3}$ for $\a=0.5$, while the smallest gaps we could analyse
using our best numerical data were around $z\approx0.05$.  Clearly, under these conditions one cannot
expect to measure the predicted $\beta=\alpha$. 

Ultimately, however, the question remains largely open. It is possible, in principle, that when the true
correlations are taken into account there exists a fairly wide region of $z>z_*$ where the local slope
$g(z)\approx\beta>\a$ is nearly constant.  In that case, the region $0<z<z_*$ would be analogous
to a boundary layer within which the mean-field result prevails and beyond which a different behaviour sets
in.

\ack
The authors thank financial support from the NSF. PP is grateful
for the warm hospitality while visiting Potsdam during this project,
and also acknowledges financial support from MIUR (PRIN 2007JHLPEZ).

\end{document}